\documentclass{pasj01}

\usepackage{lineno}

\begin{document} 
\Received{}
\Accepted{}

\title{Spatially-resolved relation between [C \emissiontype{I}] $^{3}P_{1}$--$^{3}P_{0}$ and $^{12}$CO~(1--0) in Arp~220}

\author{Junko Ueda\altaffilmark{1}}
\author{Tomonari Michiyama\altaffilmark{1,2}}
\author{Daisuke Iono\altaffilmark{1,3}}
\author{Yusuke Miyamoto\altaffilmark{1}}
\author{Toshiki Saito\altaffilmark{1,4}}

\altaffiltext{1}{
National Astronomical Observatory of Japan, National Institutes of Natural Sciences, 
2-21-1 Osawa, Mitaka, Tokyo, 181-8588}
\email{junko.ueda@nao.ac.jp}
\altaffiltext{2}{
Department of Earth and Space Science, Graduate School of Science, Osaka University, 
1-1 Machikaneyama, Toyonaka, Osaka 560-0043, Japan}
\altaffiltext{3}{
Department of Astronomical Science, The Graduate University for Advanced Studies, SOKENDAI, 
2-21-1 Osawa, Mitaka, Tokyo 181-8588, Japan}
\altaffiltext{4}{
College of Engineering, Nihon University, 
1 Nakagawara, Tokusada, Tamuramachi, Koriyama, Fukushima 963-8642, Japan}


\maketitle

\begin{abstract}
We present $\sim$0\farcs3 (114~pc) resolution maps of [C \emissiontype{I}] $^{3}P_{1}$--$^{3}P_{0}$ 
(hereafter [C \emissiontype{I}]~(1--0)) and $^{12}$CO~(1--0) obtained toward Arp~220 
with the Atacama Large Millimeter/submillimeter Array. 
The overall distribution of the [C \emissiontype{I}]~(1--0) emission is consistent with the CO~(1--0). 
While the [C \emissiontype{I}]~(1--0) and CO~(1--0) luminosities of the system follow 
the empirical linear relation for the unresolved ULIRG sample, 
we find a sublinear relation between [C \emissiontype{I}]~(1--0) and CO~(1--0) using the spatially-resolved data. 
We measure the [C \emissiontype{I}]~(1--0)/CO~(1--0) luminosity ratio per pixel in star-forming environments of Arp~220 
and investigate its dependence on the CO~(3--2)/CO~(1--0) ratio ($R_{\rm CO}$). 
On average, the [C \emissiontype{I}]~(1--0)/CO~(1--0) luminosity ratio is almost constant up to $R_{\rm CO} \simeq 1$ 
and then increases with $R_{\rm CO}$. 
According to the radiative transfer analysis, a high C \emissiontype{I}/CO abundance ratio is required 
in regions with high [C \emissiontype{I}]~(1--0)/CO~(1--0) luminosity ratios and $R_{\rm CO} > 1$, 
suggesting that the C \emissiontype{I}/CO abundance ratio varies at $\sim$100~pc scale in Arp~220. 
The [C \emissiontype{I}]~(1--0)/CO~(1--0) luminosity ratio depends on multiple factors and may not be straightforward to interpret. 
We also find the high-velocity components traced by [C \emissiontype{I}]~(1--0) in the western nucleus, 
likely associated with the molecular outflow. 
The [C \emissiontype{I}]~(1--0)/CO~(1--0) luminosity ratio in the putative outflow is 0.87 $\pm$ 0.28, 
which is four times higher than the average ratio of Arp~220. 
While there is a possibility that the [C \emissiontype{I}]~(1--0) and CO~(1--0) emission traces different components, 
we suggest that the high line ratios are likely because of elevated C \emissiontype{I}/CO abundance ratios 
based on our radiative transfer analysis. 
A C \emissiontype{I}-rich and CO-poor gas phase in outflows could be caused 
by the irradiation of the cosmic rays, the shock heating, and the intense radiation field.
\end{abstract}


\section{Introduction}

Cold molecular gas is the fuel for star formation that is a crucial process to drive the evolution of galaxies. 
The lowest rotational transition of carbon monoxide ($^{12}$CO~(1--0), hereafter CO~(1--0)) is commonly used 
as a tracer of molecular gas mass in nearby galaxies (e.g., \cite{Kennicutt1998, Saintonge2017}). 
The H$_{2}$ column density can be calculated from the luminosity of the CO~(1--0) line via the CO-to-H$_{2}$ conversion factor. 
However, the ability of CO in tracing molecular gas could be limited due to several issues, 
such as the dependence of CO-to-H$_{2}$ conversion factor on metallicity 
and gas density and the fact that CO is generally optically thick (e.g., \cite{Bolatto2013}). 

It has been proposed that one of the fine structure lines of atomic carbon, 
[C \emissiontype{I}] $^{3}P_{1}$--$^{3}P_{0}$ (hereafter [C \emissiontype{I}]~(1--0)), 
is an alternative tracer of molecular gas mass as a substitute for CO \citep{Papadopoulos2004b}. 
\citet{Papadopoulos2004a} have conducted the [C \emissiontype{I}]~(1--0) observations 
of the two typical ultraluminous infrared galaxies (ULIRGs; $10^{12}\,L_{} \leq L_{\rm IR} < 10^{13}\,L_{\solar}$), 
NGC~6240 and Arp~220, with the James Clerk Maxwell Telescope (JCMT). 
They found that the molecular gas mass estimated from the [C \emissiontype{I}]~(1--0) data 
agrees with the mass calculated from the CO~(1--0) data. 
\citet{Jiao2019} found that the [C \emissiontype{I}]~(1--0) luminosity 
correlates linearly with the CO~(1--0) luminosity for the different types of galaxies (see also \cite{Jiao2017}). 
However, recent studies claim that caution is needed 
when the [C \emissiontype{I}]~(1--0) line is used as a tracer of molecular gas mass in resolved galaxies 
because the physical conditions strongly affect the intensities (e.g., \cite{Salak2019, Saito2020, Miyamoto2021}). 
It is also predicted that the relative abundance of C \emissiontype{I} is enhanced 
in outflows and cosmic-ray (CR) dominated regions by destroying CO 
\citep{Bialy2015, Bisbas2015, Papadopoulos2018}. 
Therefore, it is necessary to understand how the [C \emissiontype{I}]~(1--0) intensity 
and its relation to the CO~(1--0) change in various regions of galaxies. 
In this study, we investigate the relationship between [C \emissiontype{I}]~(1--0) and CO~(1--0) 
at $\sim$100~pc scale in Arp~220, 
which is a well-studied source because of its relative proximity, extreme environment and brightness. 
The availability of a wealth of auxiliary data helps us investigate the physical properties 
derived from the C \emissiontype{I} data comprehensively. 
Although Arp~220 does not represent the majority of galaxies in the local universe, 
it can be a good template for high-redshift galaxies. 
This paper presents the first high-resolution [C \emissiontype{I}] observations of Arp~220.

Arp 220 is the closest ULIRG ($L_{\rm IR} = 10^{12.28}\,L_{\solar}$; \cite{Armus2009}). 
The luminosity distance is 81~Mpc based on the parameters 
from the Planck 2015 results \citep{Planck2016} 
so that 1\arcsec\,corresponds to 380~pc. 
Arp~220 is a late-stage merger, 
but it still has a double-nucleus separated by $\sim$1\arcsec\,on the sky \citep{Noris1988}. 
Infrared observations indicate that its energy output is dominated 
by a massive starburst, probably triggered by the merging process (e.g., \cite{Genzel1998, Armus2007}). 
The presence of an active galactic nucleus (AGN) in one or both nuclei 
has been inferred from multi-wavelength observations. 
The excess of $\gamma$-ray flux indicates the presence of an AGN, providing the extra CRs, 
likely in the western nucleus \citep{Yoast-Hull2017}. 
The X-ray emission obtained with \textit{NuSTAR} appears 
to be consistent with only a starburst \citep{Teng2015}, 
but there is a possibility that a deeply buried AGN is present in this system.

The distribution and the characteristics of molecular gas have been investigated in detail 
using millimeter/submillimeter telescopes (e.g., \cite{Sakamoto2008, Greve2009, Scoville2017}). 
\citet{Sakamoto1999} found that both nuclei have nuclear gas disks 
embedded in the outer gas disk rotating around the dynamical center of the system. 
The [C \emissiontype{I}] observations were carried out using the JCMT \citep{Papadopoulos2004a}, 
the \textit{Herschel} \citep{Rangwala2011, Israel2015, Kamenetzky2016}, 
and the Atacama Compact Array \citep{Michiyama2021}, 
but the spatial resolutions are larger than 1~kpc.

This paper is organized as follows. 
We summarize the archival ALMA data of Arp~220 and the data reduction procedure in Section~2. 
We show new results in Section~3. 
In Section~4, we discuss the relation between the [C \emissiontype{I}]~(1--0) and CO~(1--0) luminosities 
and the dependence of the [C \emissiontype{I}]~(1--0)/CO~(1--0) luminosity ratio on the CO~(3--2)/CO~(1--0) ratio. 
We summarize this paper in Section~5.

\section{Archival ALMA data}
We used the archival ALMA data of Arp~220, which were obtained with the 12~m array. 
Data calibration and imaging were carried out 
using the Common Astronomy Software Applications package (CASA; \cite{McMullin2007}). 
We used the appropriate versions for calibration and CASA 5.6.1 for imaging.

\subsection{[C \emissiontype{I}]~(1--0)}
We used the [C \emissiontype{I}]~(1--0) ($\nu_{\rm rest}$ = 492.161~GHz) data 
taken as a part of 2013.1.00368S. 
One spectral window was set to cover the redshifted [C \emissiontype{I}]~(1--0) line. 
The bandwidth of the spectral window is 1.875~GHz, and the frequency resolution is 1.953 MHz. 
The system temperatures ranged from 600~K to 1000~K at the observing frequency. 
The data were obtained using single pointing. 
The primary beam of the 12 m antenna is 12\arcsec. 
The quasar J1256-0547 was observed for bandpass calibration, 
and the quasar J1516+1932 was observed for phase calibration. 
Absolute flux calibration was performed using Titan. 
We adopt the typical systematic errors on the absolute flux calibration 
of 20\% for the Band~8 data \citep{THCy2}.

We first restored the calibrated measurement set 
using the observatory-provided reduction script (CASA version 4.3.1). 
We added flagging commands into the script to improve the calibration. 
After the calibration, the number of antennas available is 35. 
We made an image of the continuum emission 
using channels with negligible contamination from the spectral line. 
Next, we created a gain table of phase-only self-calibration using the continuum map. 
Then, we applied the gain table of self-calibration to all the data, including the line data. 
After the continuum subtraction, we created a data cube with a velocity resolution of 5~km~s$^{-1}$. 
As a result of self-calibration, 
the peak signal-to-noise ratios increase, and the noise levels decrease 
in channels where the bright emission were detected. 
The size of the synthesized beam is 0\farcs298 $\times$ 0\farcs219 (pa = -27\fdg7) 
by adopting Briggs weighting of the visibilities (robust= 0.5). 
The image rms per channel is 6.3 mJy beam$^{-1}$.

\subsection{$^{12}$CO~(1--0)}
We used the CO~(1--0) ($\nu_{\rm rest}$ = 115.271~GHz) data taken as a part of 2017.1.00042.S. 
Although there are three scheduling blocks (SBs) targeting the CO~(1--0) emission in the project, 
we used two SBs whose $uv$ range matches the [C \emissiontype{I}] data. 
One spectral window was set to cover the redshifted CO~(1--0) line. 
The bandwidth of the spectral window is 1.875~GHz, and the frequency resolution is 1.953~MHz. 
The system temperatures ranged from 70~K to 200~K at the observing frequency. 
The data were obtained using single pointing. 
The primary beam of the 12-m antenna is 51\arcsec. 
The quasar J1550+0527 was observed for amplitude and bandpass calibration, 
and the quasar J1532+2344 was observed for phase calibration. 
We adopt the typical systematic errors on the absolute flux calibration 
of 5\% for the Band~3 data \citep{PGCy5}.

The data reduction procedure was the same as that for the [C \emissiontype{I}] data. 
The calibration was done by using the CASA version 5.1.1. 
After the self-calibration and continuum subtraction, 
we clipped the visibilities to have the shortest $uv$ range 
(the $uv$ distance $\simeq$ 26.9~k$\lambda$) similar to the [C \emissiontype{I}] data. 
Then we created a data cube with a velocity resolution of 5~km~s$^{-1}$. 
The size of the synthesized beam is 0\farcs238 $\times$ 0\farcs200 (P.A. = -12\fdg8) 
by adopting Briggs weighting of the visibilities (robust = 0.5). 
The image RMS per channel is 0.70 mJy beam$^{-1}$.

\subsection{Supplemental data: $^{12}$CO~(3--2)}
We also used the CO~(3--2) ($\nu_{\rm rest}$ = 345.796~GHz) data 
taken as a part of 2015.1.00113.S in the discussion. 
One spectral window was set to cover the redshifted CO~(3--2) line. 
The bandwidth of the spectral window is 1.875~GHz, and the frequency resolution is 3.906~MHz. 
The system temperatures ranged from 80~K to 250~K at the observing frequency. 
The data were obtained using a single pointing with a primary beam of 17\arcsec. 
The quasars J1751+0939, J1550+0527, and J1516+1932 were observed 
for bandpass, amplitude, and phase calibration, respectively. 
We adopt the typical systematic errors on the absolute flux calibration 
of 10\% for the Band~7 data \citep{PGCy3}.

The data reduction procedure was the same as that for the [C \emissiontype{I}] data. 
The calibration was done by using the CASA version 4.7.2. 
After the self-calibration and continuum subtraction, 
we clipped the visibilities to have the shortest $uv$ range 
(the $uv$ distance $\simeq$ 26.9~k$\lambda$) similar to the [C \emissiontype{I}] data. 
We created a data cube with a velocity resolution of 5~km~s$^{-1}$. 
The size of the synthesized beam is 0\farcs239 $\times$ 0\farcs158 (P.A. = -33\fdg3) 
by adopting Briggs weighting of the visibilities (robust = 0.5). 
The image RMS per channel is 1.42 mJy beam$^{-1}$. 
We note that the CO~(3--2) spectra are affected by contamination from other molecular lines, 
possibly H$^{13}$CN~(4--3) ($\nu_{\rm rest}$ = 345.340~GHz) \citep{Wheeler2020}, 
in the velocity range of $>$5630 km~s$^{-1}$. 
We use only the CO~(3--2) data in the velocity range of 4845--5630 km~s$^{-1}$ in \S 4.2.

\section{Results}
\subsection{Distribution and Kinematics}
\begin{figure}[htbp]
    \begin{center}
        \includegraphics[width=\textwidth]{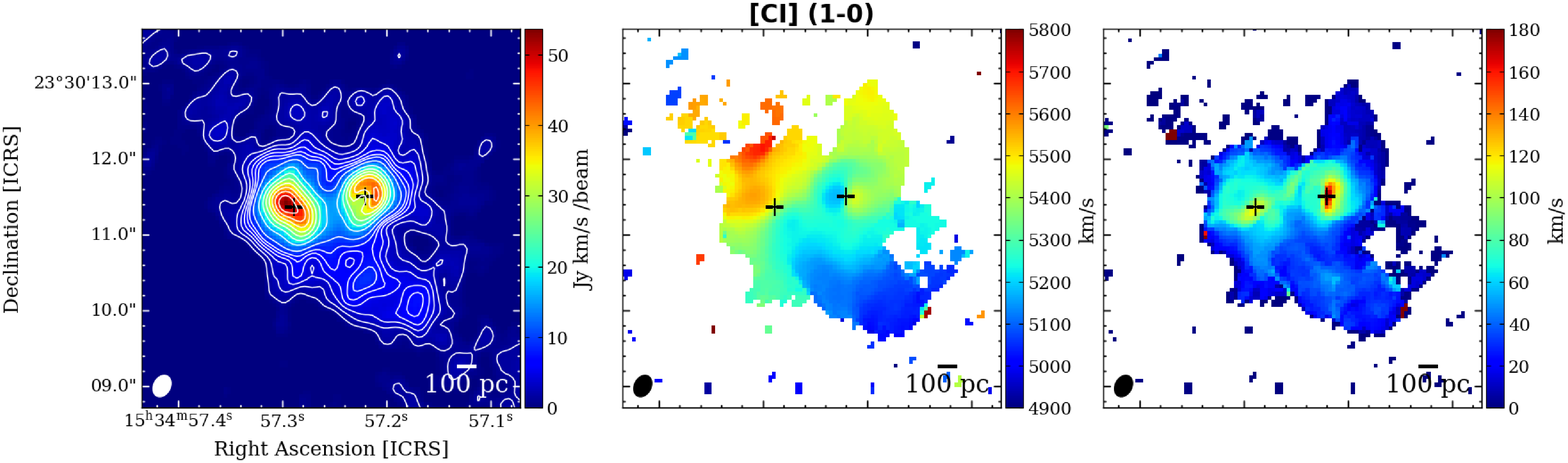}
        \includegraphics[width=\textwidth]{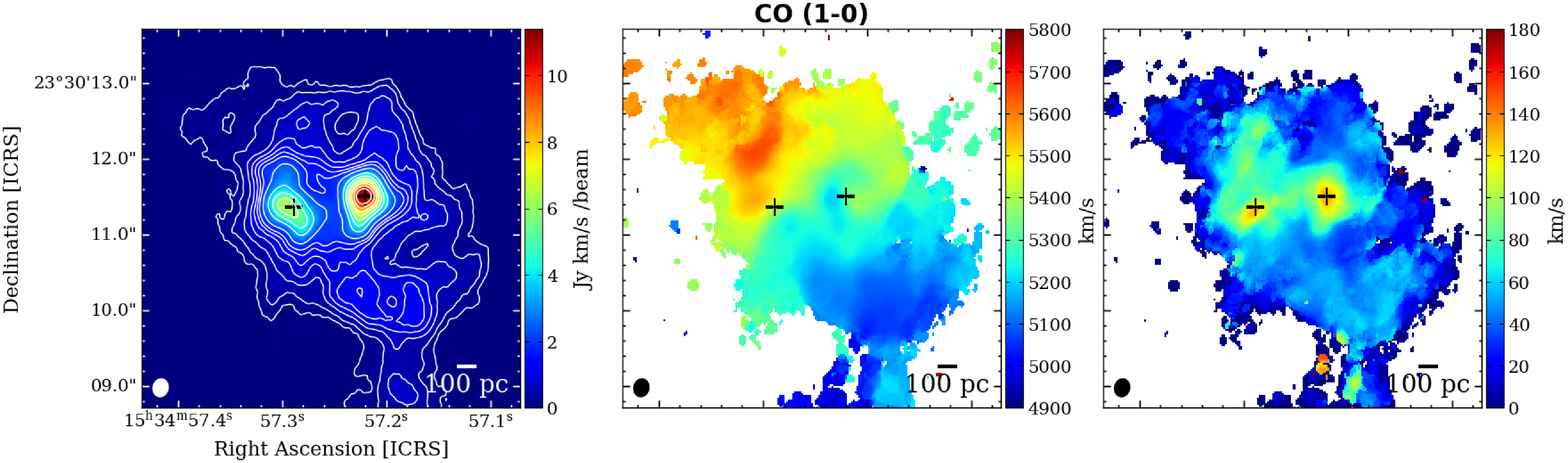}
    \end{center}
    \caption{
    (top) The [C \emissiontype{I}]~(1--0) integrated intensity, 
    velocity field, and velocity dispersion maps of Arp~220. 
    The integrated velocity ranges are 1050 km~s$^{-1}$ (4845--5895 km~s$^{-1}$).
    The contour levels of the integrated intensity map are the peak integrated intensity 
    (54 Jy~km~s$^{-1}$ beam$^{-1}$) $\times$ 
    (0.025, 0.050, 0.075, 0.100, 0.125, 0.150, 0.2, 0.3, 0.4, 0.5, 0.6, 0.7, 0.8, 0.9).
    The plus signs show the nuclei defined by the 880 $\mu$m continuum emission, 
    and the ellipse in the bottom left corner shows the beam size.
    (bottom) Same as the top figures but for the CO~(1--0).
    The contour levels of the integrated intensity map are the peak integrated intensity 
    (11 Jy~km~s$^{-1}$ beam$^{-1}$) $\times$ 
    (0.010, 0.025, 0.050, 0.075, 0.100, 0.125, 0.150, 0.2, 0.3, 0.4, 0.5, 0.6, 0.7, 0.8, 0.9).
\label{fig:f1}}
\end{figure}

\begin{figure}[htbp]
    \begin{center}
        \includegraphics[width=0.5\textwidth]{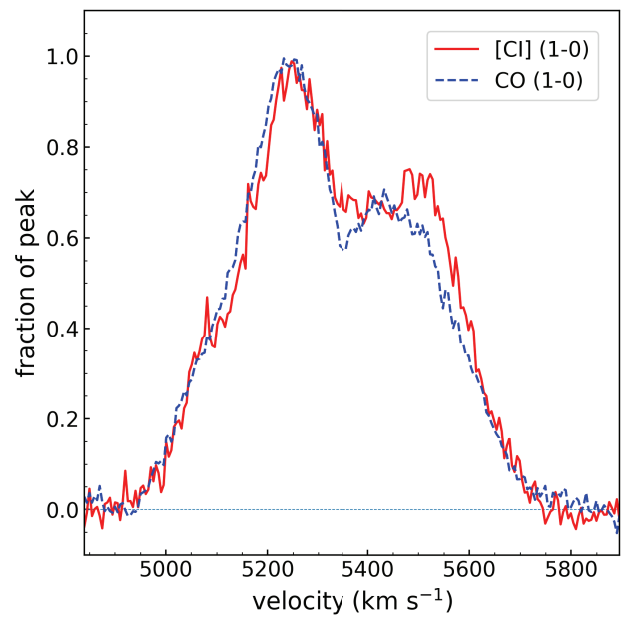}
    \end{center}
    \caption{
    The [C \emissiontype{I}]~(1--0) and CO~(1--0) spectra for the central 5\arcsec region 
    are presented by the solid red and dashed blue lines, respectively.
    The y-axis is the flux density normalized by the peak.
\label{fig:f2}}
\end{figure}

The integrated intensity, velocity field, and velocity dispersion maps 
of [C \emissiontype{I}]~(1--0) and CO~(1--0) are presented in Figure~\ref{fig:f1}. 
The integrated intensity maps were created without clipping the intensities. 
The velocity field and velocity dispersion maps were made 
after clipping the cleaned image cubes at the 4$\sigma$ level per channel. 
The apparent distribution of the [C \emissiontype{I}]~(1--0) emission 
is similar to the CO~(1--0) distribution overall. 
However, the strongest [C \emissiontype{I}]~(1--0) and CO~(1--0) peaks are inconsistent, 
and they are located in the eastern and western nuclei, respectively. 
An arc-like feature in the [C \emissiontype{I}]~(1--0) map seen around the western nucleus 
is due to self-absorption. 
The [C I]~(1--0) velocity dispersion is 186~km~s$^{-1}$ and 117~km~s$^{-1}$ 
in the western and eastern nuclei, respectively. 
On the other hand, the CO~(1--0) velocity dispersion is 
133~km~s$^{-1}$ and 128~km~s$^{-1}$ in the western and eastern nuclei, respectively. 
Thus, the [C \emissiontype{I}]~(1--0) velocity dispersion is 
$\sim$1.4 times higher than the CO~(1--0) in the western nucleus. 
In addition, we present the [C \emissiontype{I}]~(1--0) and CO~(1--0) spectra 
for the central 5\arcsec region (the region with a radius of 2\farcs5 from the pointing center 
located between the two nuclei) in Figure~\ref{fig:f2}. 
The spectral profiles are similar, but subtle differences can be seen at $v$ = 5100--5600 km~s$^{-1}$.

\subsection{Molecular mass}
The integrated [C \emissiontype{I}]~(1--0) line flux density 
within the central 5\arcsec (1.9~kpc) region is 760 $\pm$ 150 Jy~km~s$^{-1}$. 
The integrated CO~(1--0) line flux density within the same region is 190 $\pm$ 10 Jy~km~s$^{-1}$. 
Since our data were not corrected with the zero-spacing information and visibilities are clipped, 
the data suffer from missing flux. 
For example, we compare the integrated [C \emissiontype{I}]~(1--0) line flux density 
with the measurement taken by the JCMT 
(1160 $\pm$ 350 Jy~km~s$^{-1}$ ($\Omega_{\rm PB}$ = 10\arcsec); \cite{Papadopoulos2004a}). 
The recovered flux is 66~\%. 
The shortest $uv$ range of our data is 26.9~k$\lambda$, 
which corresponds to the maximum recoverable scale of 1.75~kpc for Arp~220. 
Thus, we discuss components smaller than 1.75~kpc in this study. 
We calculate the line luminosity from the integrated line flux density using Equation~(3) of \citet{Solomon2005}:
\begin{equation}
L'_{\rm line} = 3.25 \times 10^{7} S_{\rm line}\Delta\,v \nu_{\rm obs}^{-2} D_{\rm L}^{2} (1 + z)^{-3},
\end{equation}
where $L'_{\rm line}$ is the luminosity in K km s$^{-1}$ pc$^{2}$, 
$S_{\rm line}\Delta\,v$ is the integrated line flux density in Jy km s$^{-1}$, 
$\nu_{\rm obs}$ is the observing frequency in GHz, 
$D_{\rm L}$ is the luminosity distance in Mpc, 
and $z$ is the redshift. 
The derived line luminosity is summarized in Table~\ref{tab:t1}. 
We calculate the [C \emissiontype{I}]~(1--0)/CO~(1--0) 
(hereafter [C \emissiontype{I}]/CO) luminosity ratio of Arp~220. 
The derived luminosity ratio is 0.22 $\pm$ 0.04, which is comparable 
to the average [C \emissiontype{I}]/CO ratio (0.22 $\pm$ 0.09) of 20 U/LIRGs 
presented in \citet{Valentino2018}.

Assuming the optically thin condition, 
we calculate the molecular gas mass ($M_{\rm CI}$) 
from the integrated [C \emissiontype{I}]~(1--0) line flux density 
using Equation (6) of \citet{Bothwell2017}:
\begin{equation}
M_{\rm CI}({\rm H_{2}}) = 1375.8 \frac{D_{\rm L}^2}{1+z}\left(\frac{X_{\rm CI}}{10^{-5}}\right)^{-1}\left(\frac{A_{10}}{10^{-7} {\rm s^{-1}}}\right)^{-1}Q_{10}^{-1}S_{\rm CI}\Delta v,
\end{equation}
where $M_{\rm CI}$(H$_{2}$) is the molecular gas mass in the unit of $M_{\solar}$, 
$X_{\rm CI}$ is the C \emissiontype{I} abundance relative to H$_{2}$, 
$A_{10}$ is the Einstein coefficient (= $7.93 \times 10^{-8}$~s$^{-1}$), 
$Q_{10}$ is the [C \emissiontype{I}] excitation factor, 
and $S_{\rm CI}\Delta v$ is the integrated [C \emissiontype{I}]~(1--0) line flux density. 
We calculate $M_{\rm CI}$ by applying $X_{\rm CI}$ = $3 \times 10^{-5}$, 
which was used in the previous [C \emissiontype{I}]~(1--0) study of Arp 220 \citep{Papadopoulos2004a}. 
We also adopt $Q_{10}$ = 0.48, which is the mean value of $Q_{10}$ 
for the typical ISM conditions in galaxies where the [C \emissiontype{I}] lines are globally subthermally excited 
($n_{\rm H_{2}}$ = (0.3 -- 1.0) $\times$ 10$^{4}$ cm$^{-3}$ and $T_{\rm kin}$ = 25 -- 80~K) \citep{Papadopoulos2021}. 
The derived $M_{\rm CI}$ is $(6.0 \pm 1.2) \times 10^{9} M_{\solar}$. 
In addition, we calculate the molecular gas mass ($M_{\rm CO}$) from the integrated CO~(1--0) line flux density 
by using a ULIRG-like CO luminosity-to-H2 mass conversion factor 
($\alpha_{\rm CO}$ = 0.8 $M_{\solar}$ pc$^{-2}$ (K~km~s$^{-1}$)$^{-1}$; \cite{Bolatto2013}) 
and derive $M_{\rm CO}$ of $(2.4 \pm 0.1) \times 10^{9} M_{\solar}$. 
In this case, $M_{\rm CI}$ is $\sim$2.5 times larger than $M_{\rm CO}$. 
The relative abundance of CI varies from $1.6 \times 10^{-5}$ to $8.4 \times 10^{-5}$ 
depending on the source (e.g., \cite{Walter2011, Valentino2018, Boogaard2020}). 
\citet{Jiao2019} suggest that $X_{\rm CI}$ in local U/LIRGs (Avg. $X_{\rm CI} = (8.3 \pm 3.0) \times 10^{-5}$) 
is about three times higher than $X_{\rm CI}$ of spiral galaxies. 
When applying this abundance, $M_{\rm CI}$ of Arp 220 results in $(2.2 \pm 0.4) \times 10^{9} M_{\solar}$, 
which is similar to $M_{\rm CO}$. 
Assuming that $M_{\rm CI}$ is consistent with $M_{\rm CO}$ and 
the assumed $\alpha_{\rm CO}$ and $Q_{\rm 10}$ are correct, 
we derive $X_{\rm CI}$ of $7.5 \times 10^{-5}$, 
which is consistent with the average $X_{\rm CI}$ in local U/LILRGs \citep{Jiao2019} within the errors. 
Thus, the [C \emissiontype{I}]~(1--0) line can be used as 
a substitute for CO~(1--0) to estimate the molecular gas mass, 
but we need to carefully choose the parameters, 
including the relative abundance of C \emissiontype{I}.

\subsection{High-velocity components in the western nucleus}
\begin{figure}[htbp]
    \begin{center}
        \includegraphics[width=\textwidth]{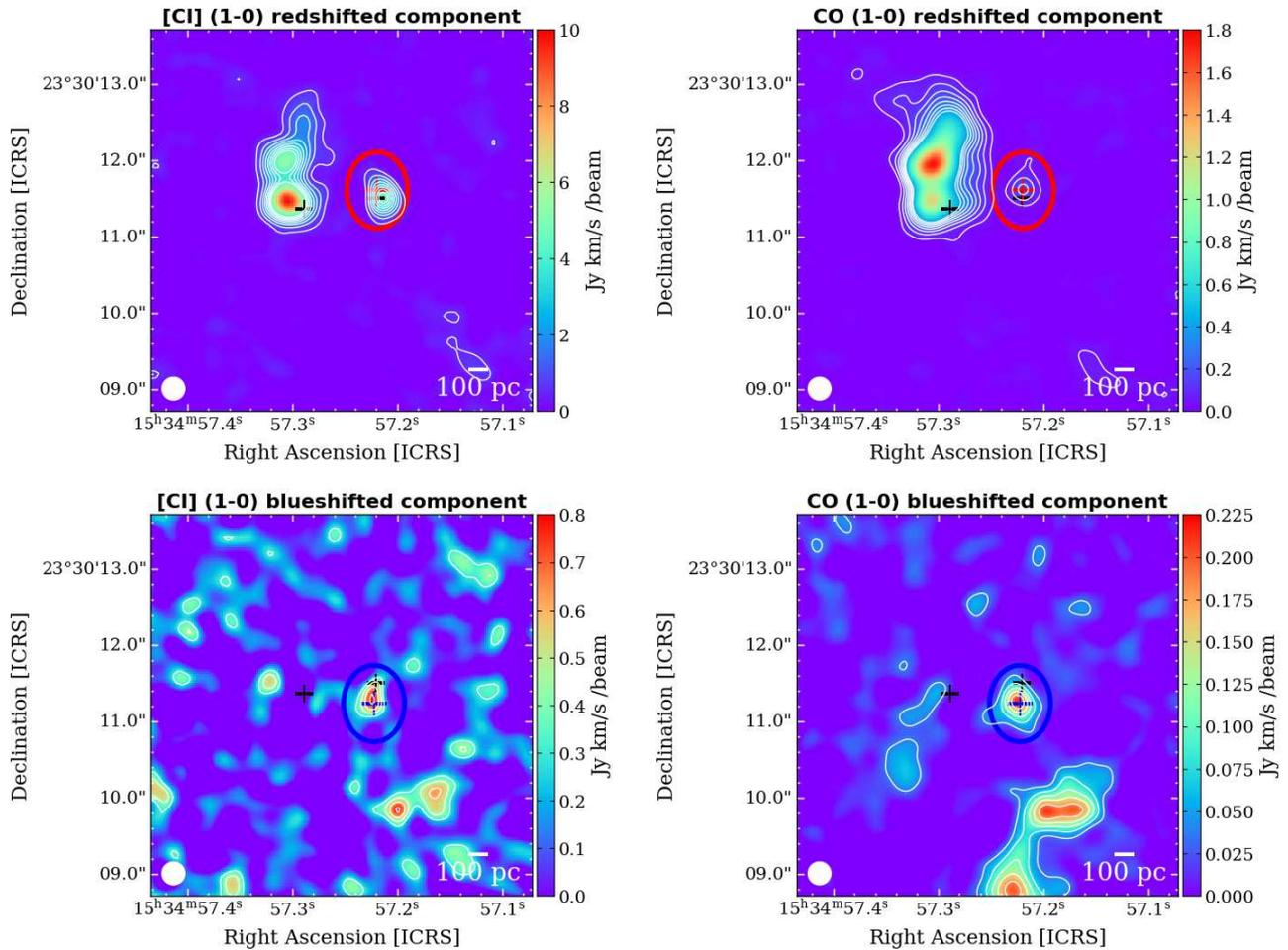}
    \end{center}
    \caption{
    (top) The [C \emissiontype{I}]~(1--0) and CO~(1--0) integrated intensity maps of the redshifted component. 
    The integrated velocity range is [5625, 5895] km~s$^{-1}$
    ($\Delta v$ = 270~km~s$^{-1}$). The contour levels are 
    0.26 Jy~beam$^{-1}$ km~s$^{-1}$ $\times$ (2, 4, 6, 8, 10, 12, 14, 16) for the [C \emissiontype{I}] map 
    and 25 mJy~beam$^{-1}$ km~s$^{-1}$ $\times$ (2, 4, 6, 8, 10, 12, 14, 16) for the CO map.
    The black plus signs show the nuclei defined by the 880 $\mu$m continuum emission, 
    and the ellipse in the bottom left corner shows the beam size ($\theta$ = 0\farcs3).
    The red plus sign shows the peak of the CO redshifted component.
    We measure the flux densities of the redshifted component (Table~\ref{tab:t2}) 
    by integrating the pixel values within the red ellipse (0\farcs8 $\times$ 1\farcs0).
    (bottom) Same as the top figures but for the blueshifted component.
    The integrated velocity range is [4845, 4990] km~s$^{-1}$ ($\Delta v$ = 140~km~s$^{-1}$). 
    The contour levels are 0.15 Jy~beam$^{-1}$ km~s$^{-1}$ $\times$ (2, 3, 4, 5) for the [C \emissiontype{I}] map 
    and 18 mJy~beam$^{-1}$ km~s$^{-1}$ $\times$ (2, 4, 6, 8, 10) for the CO map.
    The blue plus sign shows the peak of the CO blueshifted component.
    We measure the flux densities of the blueshifted component 
    by integrating the pixel values within the blue ellipse (0\farcs8 $\times$ 1\farcs0).
\label{fig:f3}}
\end{figure}

Recent high-resolution observations with ALMA discovered the collimated molecular outflow 
in the western nucleus of Arp~220 \citep{Wheeler2020, Barcos2018}. 
The molecular outflow was detected in diffuse and dense gas tracers, 
such as CO~(1--0) and HCN~(1--0) lines. 
We created the CO~(1--0) integrated intensity maps of high-velocity components 
(see Figure~\ref{fig:f3}) by using the same velocity ranges used in \citet{Barcos2018} 
because it is difficult to identify the morphology of the outflowing components 
due to the coarse angular resolution. 
The velocity ranges of the blueshifted and redshifted components are 
-510~km~s$^{-1}$ $< v - v_{\rm sys} <$ -370~km~s$^{-1}$ and 
270~km~s$^{-1}$ $< v - v_{\rm sys} <$ 540~km~s$^{-1}$, respectively. 
The systemic velocity is $v_{\rm sys}$ = 5355 $\pm$ 15~km~s$^{-1}$ 
for the western nucleus \citep{Sakamoto1999}. 
As discovered by the previous observations, 
high-velocity components were detected around the western nucleus (Figure~\ref{fig:f3}). 
The blueshifted component is located in the south of the western nucleus, 
whereas the redshifted component is located close to the nucleus. 
There are also disk components in the maps. 
In addition, we created the [C \emissiontype{I}]~(1--0) integrated intensity maps 
using the same velocity ranges (Figure~\ref{fig:f3}). 
While the blueshifted component around the western nucleus was only marginally detected, 
the redshifted component was clearly detected. 
The [C \emissiontype{I}]~(1--0) emission peaks are consistent 
with the CO~(1--0) emission peaks within the beam size. 
Since the location and velocity of the [C \emissiontype{I}]~(1--0) components are consistent with the CO~(1--0), 
the [C \emissiontype{I}]~(1--0) high-velocity components are likely associated with the outflow. 
Furthermore, although the angular resolutions are not high enough to reveal the detailed distributions, 
the redshifted components of [C \emissiontype{I}]~(1--0) and CO~(1--0) seem to be slightly separated, 
implying that the [C \emissiontype{I}]~(1--0) and CO~(1--0) distributions are different in the outflow. 
Future high-resolution observations are necessary to confirm this.

\begin{figure}[htbp]
    \begin{center}
    	\includegraphics[width=\textwidth]{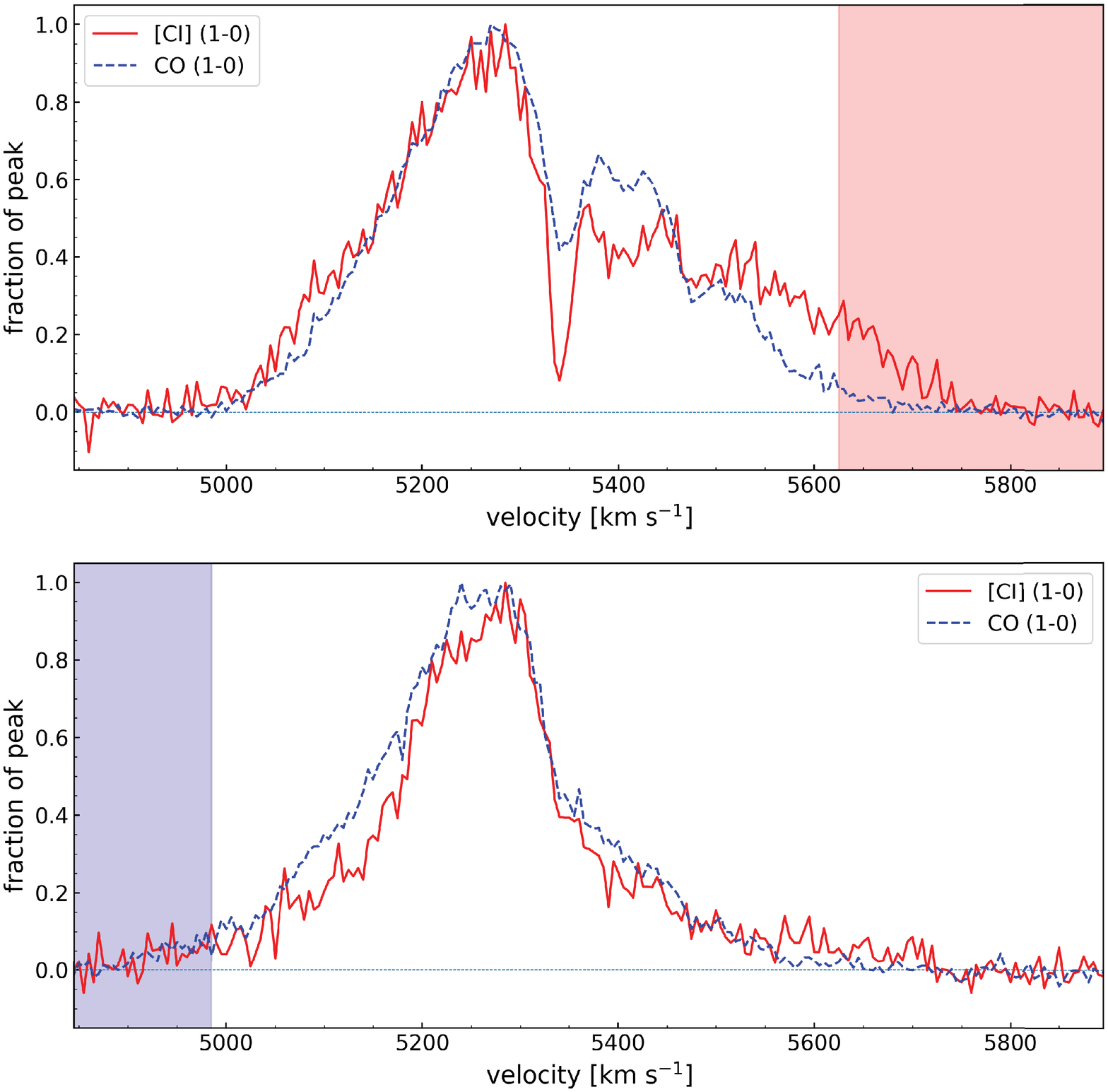}
    \end{center}
    \caption{
    (top) The [C \emissiontype{I}]~(1--0) and CO~(1--0) spectra in the peak of the CO redshifted component 
    ([\timeform{15h34m57.219s}, \timeform{+23D30'11.606''})]; see also the red plus sign in Figure~\ref{fig:f3} (top)).
    The y-axis is the flux density normalized by the peak.
    The red background color shows the velocity range of the redshifted component.
    (bottom) The spectra in the peak of the CO blueshifted component 
    ([\timeform{15h34m57.222s}, \timeform{+23D30'11.231''}]; see also the blue plus sign in Figure~\ref{fig:f3} (bottom)). 
    The blue background color shows the velocity range of the blueshifted component.
\label{fig:f4}}
\end{figure}

We calculate the molecular gas mass of the high-velocity components 
from the integrated [C \emissiontype{I}]~(1--0) and CO~(1--0) line flux densities 
in the same way as the previous section. 
Firstly, we measure the line flux densities within 0\farcs8 $\times$ 1\farcs0 
from the CO~(1--0) peak of each high-velocity component. 
The regions are shown by the blue and red ellipses in Figure~\ref{fig:f3} 
and are chosen to include the lowest (2$\sigma$) contour. 
The image RMS was measured in the line-free region of each integrated intensity map. 
Then, we calculate $M_{\rm CI}$ by applying $Q_{10}$ = 0.47 \citep{Papadopoulos2004a} 
and $X_{\rm CI} = 7.5 \times 10^{-5}$ we derive in \S 3.2 
and $M_{\rm CO}$ by using $\alpha_{\rm CO}$ = 0.8 $M_{\solar}$ pc$^{-2}$ (K~km~s$^{-1}$)$^{-1}$ \citep{Bolatto2013}. 
The derived molecular gas mass is presented in Table~\ref{tab:t2}. 
While $M_{\rm CI}$ of the blueshifted component is 0.76 times smaller than $M_{\rm CO}$, 
$M_{\rm CI}$ of the redshifted component is 5.3 times larger than $M_{\rm CO}$. 
One possibility is that $M_{\rm CI}$ of the redshifted component is overestimated 
due to the contamination of gas associated with the main body. 
The spectral profiles of the redshifted component are significantly different 
between [C \emissiontype{I}]~(1--0) and CO~(1--0) (Figure~\ref{fig:f4}), 
implying that the [C \emissiontype{I}]~(1--0) and CO~(1--0) emission traces different components. 
However, even if we take into account this possibility, 
it is not possible to explain the difference between $M_{\rm CI}$ and $M_{\rm CO}$ of the blueshifted component. 
These inconsistencies indicate that the parameters, such as $\alpha_{\rm CO}$, $X_{\rm CI}$ and $Q_{10}$, 
differ between the galaxy-averaged and outflow components 
and even between the blueshifted and redshifted components. 
For example, matching $M_{\rm CO}$ of the redshifted component with $M_{\rm CI}$ 
requires a CO luminosity-to-H$_{2}$ conversion factor of $\alpha_{\rm CO}$ = 4.3. 
This value is larger than the ULIRG-like CO luminosity-to-H2 mass conversion factor 
but similar to the standard CO luminosity-to-H2 mass conversion factor 
(e.g., $\alpha_{\rm CO}$ = 4.35 $M_{\solar}$ pc$^{-2}$ (K~km~s$^{-1}$)$^{-1}$ \cite{Bolatto2013}). 
Future high-resolution [C \emissiontype{I}] observations would reveal 
the detailed morphology of the atomic carbon outflow, 
allowing us to derive the spatial distribution of the different physical parameters.

\section{Discussion}
\subsection{The relation between [C \emissiontype{I}] and CO}
\begin{figure}[htbp]
    \begin{center}
        \includegraphics[width=0.5\textwidth]{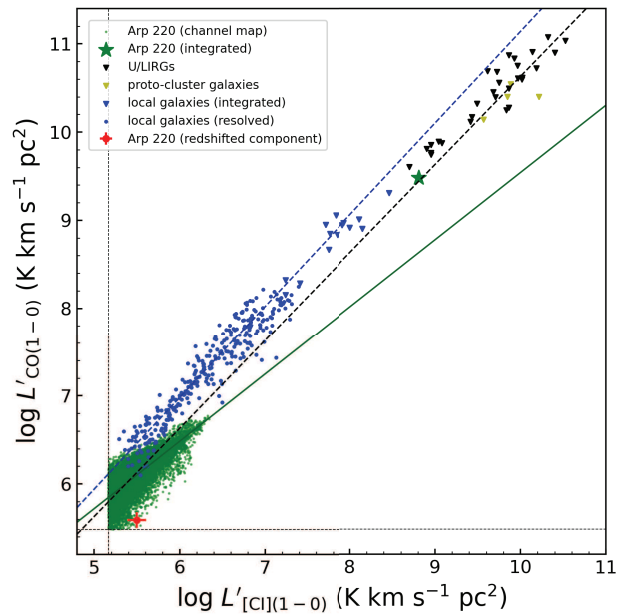}
    \end{center}
    \caption{
    Plot of the [C \emissiontype{I}]~(1--0) luminosity vs. the CO~(1--0) luminosity.
    The different symbols present the different samples.
    The green dots are star forming environments in Arp~220.
    The blue circles are resolved galaxies in the local universe \citep{Jiao2019}.
    The other symbols show the integrated values.
    The green star and red circle are the values integrated over 
    the central 5\arcsec region and redshifted component of Arp~220, respectively.
    The black triangles are U/LIRGs (\cite{Liu2015, Kamenetzky2016, Valentino2018}),
    the yellow triangles are proto-cluster galaxies at $z$ = 2,2 \citep{Emonts2018}, 
    and the blue triangles are integrated values of local galaxies \citep{Jiao2019}.
\label{fig:f5}}
\end{figure}

We plot the [C \emissiontype{I}]~(1-0) and CO~(1--0) luminosities of Arp~220 
and the literature samples of local galaxies (non-U/LIRGs) \citep{Jiao2019}, 
unresolved U/LIRGs (\cite{Liu2015, Kamenetzky2016, Valentino2018}), 
and proto-cluster galaxies at $z$ = 2.2 \citep{Emonts2018} in Figure~\ref{fig:f5}. 
\citet{Jiao2019} found two different linear relations in the [C \emissiontype{I}]~(1--0) vs. CO~(1--0) plot. 
One is for the non-U/LIRGs, and the other is for the U/LIRGs and the proto-cluster galaxies. 
The best-fit lines for these two groups are shown by dashed blue and black lines in Figure~\ref{fig:f5}. 
Arp~220 (integrated) is consistent with the linear relation for the unresolved ULIRG sample.

We check whether the [C \emissiontype{I}]~(1--0) luminosity correlates with the CO~(1--0) luminosity 
at $\sim$100~pc scale using the channel maps of Arp~220. 
We create beam-matched maps by smoothing the original channel maps and perform the pixel binning. 
The final angular resolution is 0\farcs3, which corresponds to $\sim$110~pc. 
The pixel size is 57~pc $\times$ 57~pc, and the velocity resolution is 5~km~s$^{-1}$. 
Then, we measure the [C \emissiontype{I}]~(1--0) and CO~(1--0) luminosities per pixel, but we exclude pixels 
in which either [C \emissiontype{I}]~(1--0) or CO~(1--0) flux densities are below the 3$\sigma$ level. 
The lower limits of the [C \emissiontype{I}]~(1--0) and CO~(1--0) luminosities are 
$1.5 \times 10^{5}$ K~km~s$^{-1}$~pc$^{2}$ and $3.0 \times 10^{5}$ K~km~s$^{-1}$~pc$^{2}$, respectively. 
We also exclude pixels within 300~pc from each nucleus because self-absorption can affect our interpretation.

We plot the [C \emissiontype{I}]~(1--0) and CO~(1--0) luminosities measured in each pixel 
(Arp~220 (channel map)) in Figure~\ref{fig:f5} 
and perform a linear fitting to the [C \emissiontype{I}]~(1--0) and CO~(1--0) data points 
using the python package linmix \citep{Kelly2007}. 
This gives the best-fit line, 
\begin{equation}
\log L'_{\rm CO(1-0)} = (0.76 \pm 0.01) \log L'_{\rm [CI](1-0)} + (1.89 \pm 0.05),
\end{equation}
but we exclude pixels in which the [C \emissiontype{I}]~(1--0) luminosity 
is below log\,$L'_{\rm [CI](1-0)}$ = 5.3 for fitting 
because the sampled data are dominated by noise. 
The slope of the best-fit line is smaller than the unity. 
Such a sublinear relation between $L'_{\rm [CI](1-0)}$ and $L'_{\rm CO(1-0)}$ 
has been found in the previous study on a ULIRG, IRAS F18293-3413 \citep{Saito2020}. 
While the [C \emissiontype{I}]~(1--0) and CO~(1--0) luminosities integrated across the system of Arp~220 
follow the empirical linear relation for the unresolved ULIRG sample, 
we find a sublinear relation between $L'_{\rm [CI](1-0)}$ and $L'_{\rm CO(1-0)}$ 
using the spatially-resolved data of Arp~220, 
indicating that the [C \emissiontype{I}]/CO luminosity ratios are not constant in the system.

\subsection{Variations in the C \emissiontype{I}/CO abundance ratios}
\begin{figure}[htbp]
    \begin{center}
        \includegraphics[width=0.5\textwidth]{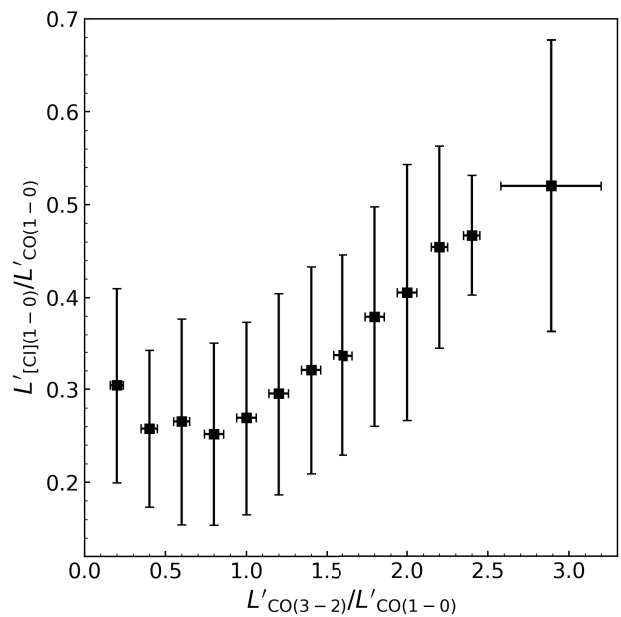}
    \end{center}
    \caption{
    Plot of the average [C \emissiontype{I}]~(1--0)/CO~(1--0) luminosity ratio vs. the CO~(3--2)/CO~(1--0) luminosity ratio.
\label{fig:f6}}
\end{figure}

We check the dependence of the [C \emissiontype{I}]/CO luminosity ratio 
on the CO~(3--2)/CO~(1--0) ratio ($R_{\rm CO}$) 
in order to investigate what causes variations of $L'_{\rm [CI](1-0)}$/$L'_{\rm CO(1-0)}$ in Arp~220. 
In general, $R_{\rm CO}$ is enhanced in active star forming regions 
because increasing the gas density and heating the interstellar medium 
due to ultraviolet (UV) emission from newly born stars 
are responsible for collisionally exciting the CO gas to the $J$ = 3 level. 
Thus, $R_{\rm CO}$ can be interpreted as an indicator of the physical properties of the gas. 
We divide the pixels into bins with $\Delta R_{\rm CO}$ = 0.2 (0.1 $\leq R_{\rm co} <$ 2.5) 
and the others ($R_{\rm co} \geq 2.5$) because the sample size is small in a range of $R_{\rm co} \geq 2.5$. 
Then, we calculate the average $L'_{\rm [CI](1-0)}/L'_{\rm CO(1-0)}$ of each bin 
and plot them as a function of $R_{\rm CO}$ in Figure~\ref{fig:f6}. 
As a result, the average $L'_{\rm [CI](1-0)}/L'_{\rm CO(1-0)}$ is roughly constant up to $R_{\rm CO}$ = 1.0 
and then increases with $R_{\rm CO}$.

We perform non-local thermodynamic equilibrium (non-LTE) analysis 
using the radiative transfer code RADEX \citep{vanderTak2007} 
to check whether this trend can be reproduced by changing the physical properties of gas. 
RADEX has been developed to infer the physical and chemical parameters, 
such as temperature, density, and molecular abundances, in molecular clouds. 
It requires the line width that reflects the internal velocity dispersions of molecular clouds 
as one of the input parameters. 
However, we cannot measure the line width using our maps due to coarse angular resolutions. 
We thus adopt the typical line-width of molecular clouds in the Galaxy 
(5~km~s$^{-1}$; e.g., \cite{Heyer2009}), assuming that similar molecular clouds form in Arp 220. 
We fixed the CO column density ($N_{\rm CO}$) to 5 $\times$ 10$^{17}$\,cm$^{-2}$ 
based on the radiative transfer analysis \citep{Sliwa2017}. 
They performed RADEX calculations to model the CO lines of Arp 220 
and derived the mean $N_{\rm co}$ of (1.1--2.3) $\times 10^{19}$ cm$^{-2}$ (d$V$ = 120--320 km~s$^{-1}$). 
This gives $N_{\rm co}$/d$V$ of (7--9) $\times$ 10$^{16}$. 
We thus adopt $N_{\rm CO}$ of $5 \times 10^{17}$ (d$V$ = 5 km~s$^{-1}$). 
We also fix the background temperature to 2.73~K. 
We vary the C \emissiontype{I} column density ($N_{\rm CI}$) 
and the kinetic temperature ($T_{\rm k}$) or the H$_{2}$ density ($n_{\rm H_{2}}$), 
and calculate the radiation temperature of the lines. 
RADEX yields the opacity of [C \emissiontype{I}]~(1--0) $<$1 
and the opacity of CO~(3--2) $>$1 for most conditions. 
The opacity of CO~(1--0) is large ($>$1) at the low $T_{\rm kin}$ and $n_{H_{2}}$ 
and decreases with increasing $T_{\rm kin}$ and $n_{\rm H_{2}}$. 
The CO~(1--0) is optically thin in regions with $R_{\rm CO} > 1$. 

\begin{figure}[htbp]
    \begin{center}
        \includegraphics[width=0.32\textwidth]{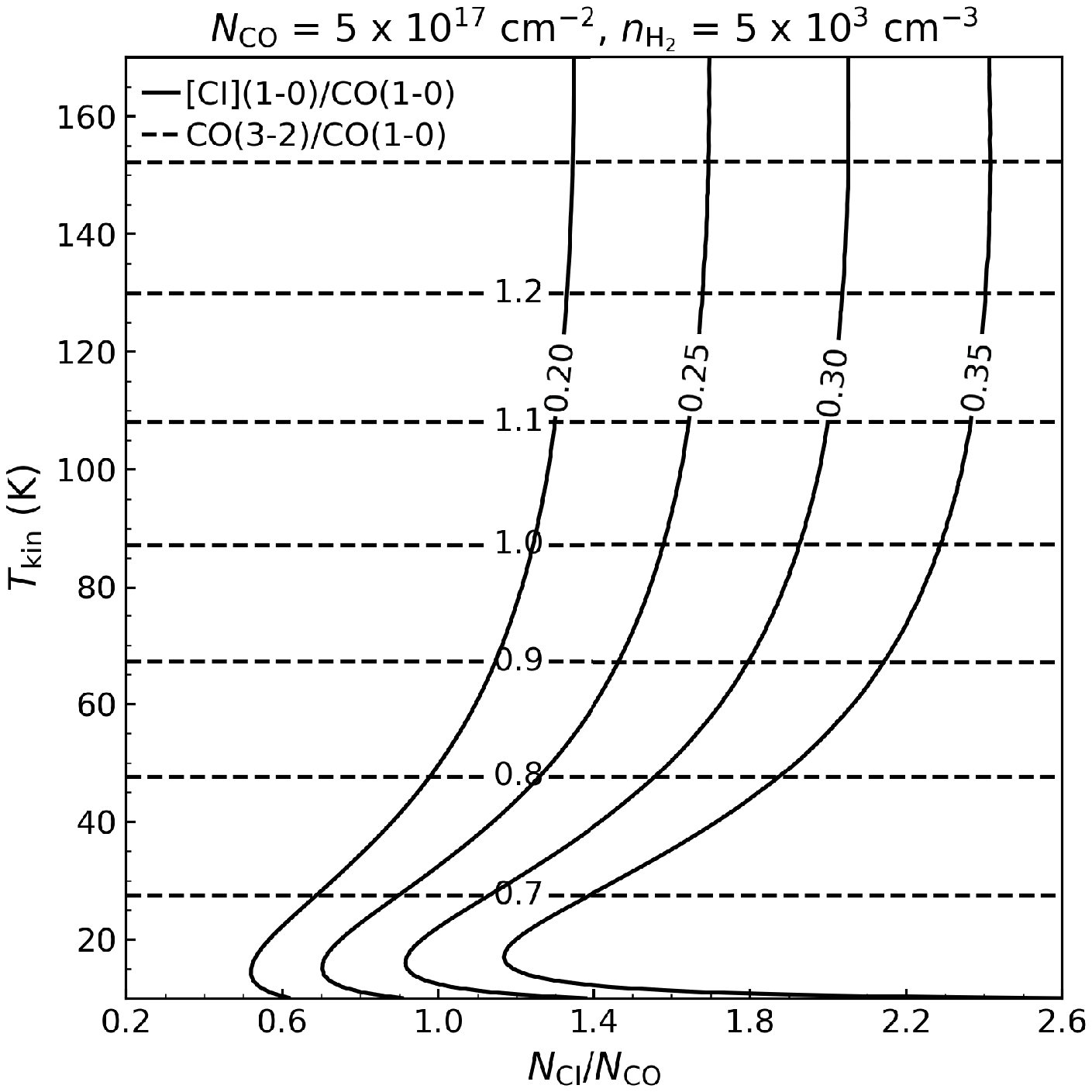}
        \includegraphics[width=0.32\textwidth]{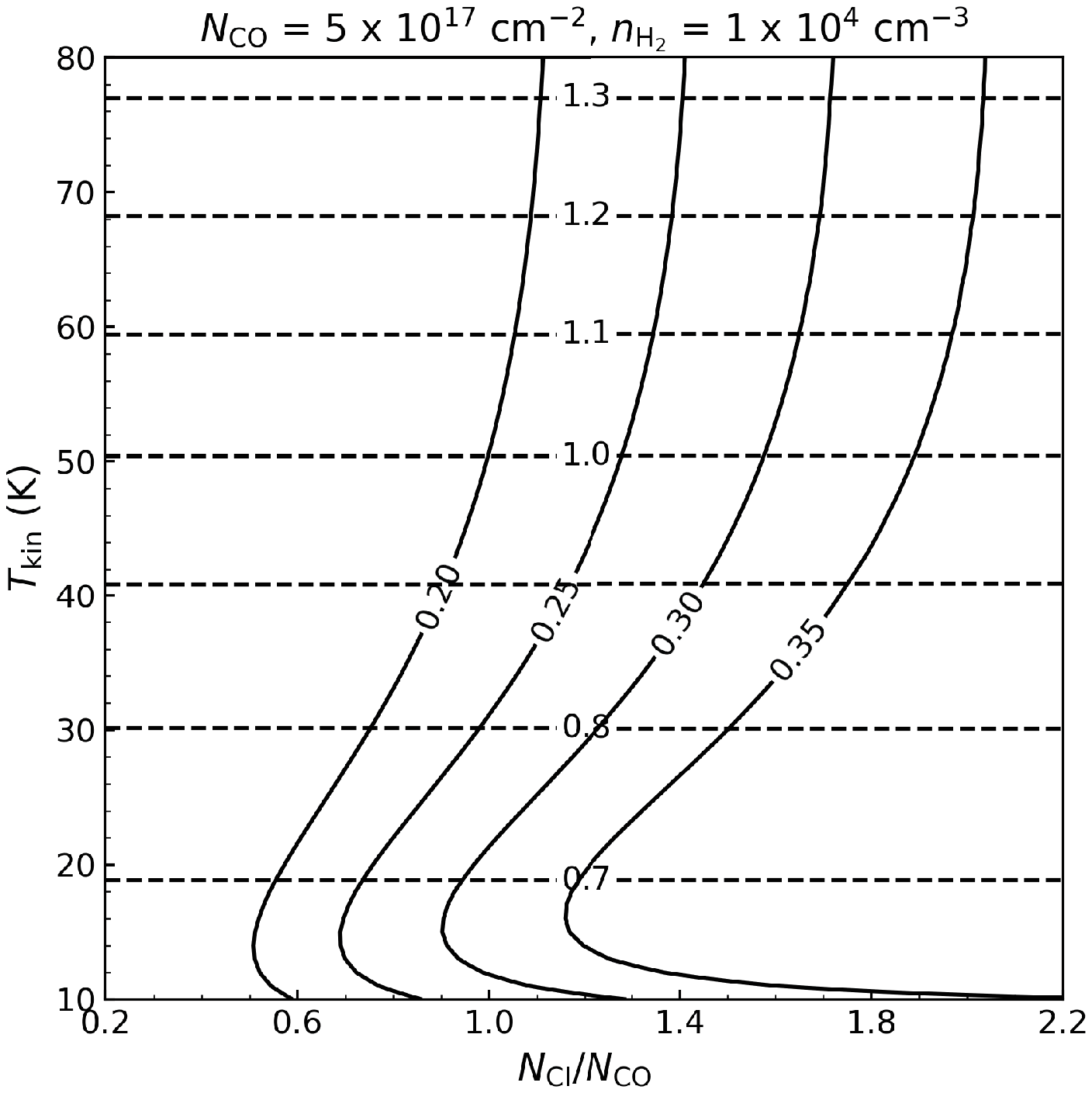}
        \includegraphics[width=0.32\textwidth]{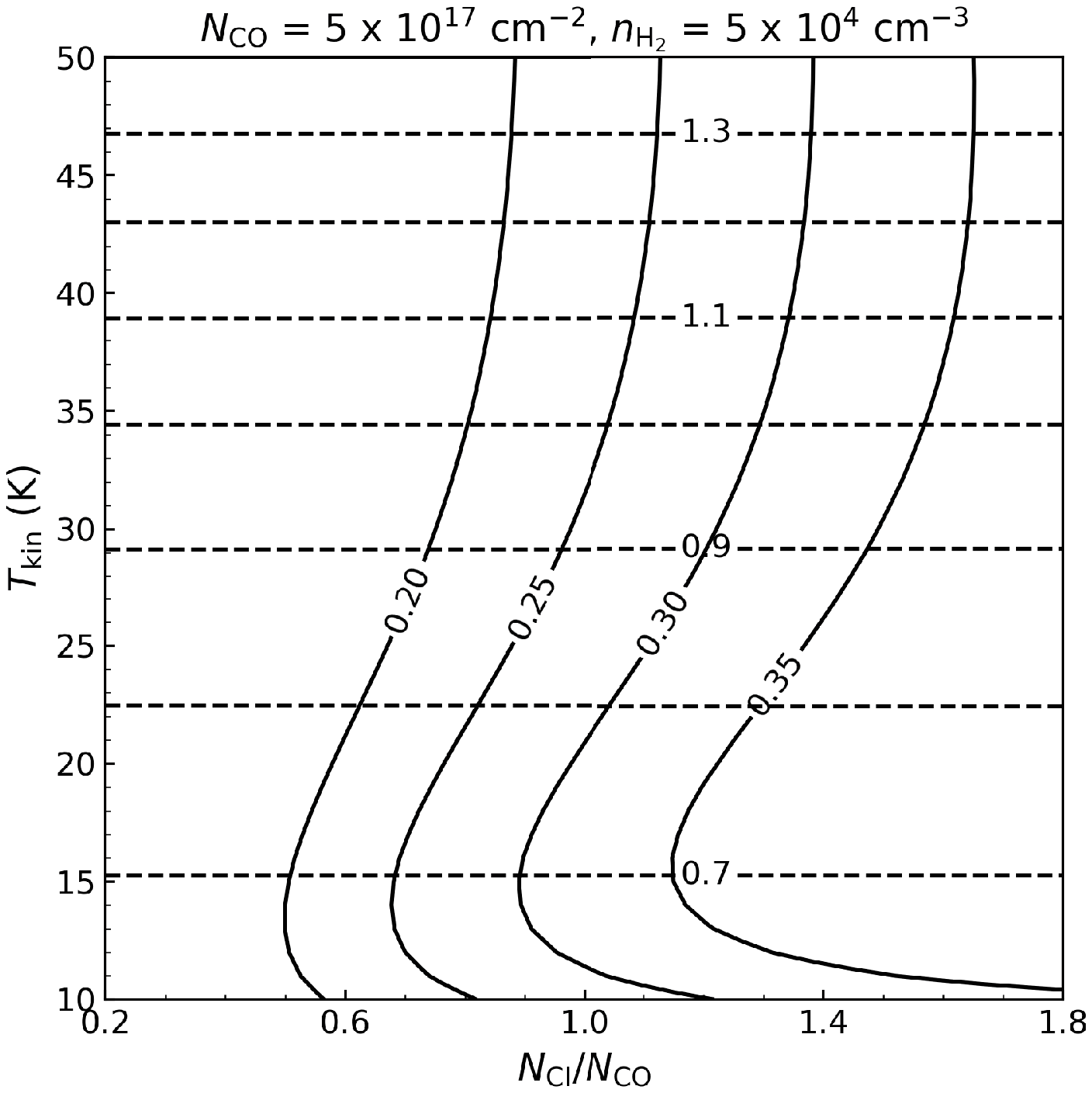}        
    \end{center}
    \caption{
    Results of RADEX calculation when the H$_{2}$ density is fixed to 
    $n_{\rm H_{2}}$ = [0.5, 1.0, 5.0] $\times$ 10$^{4}$~cm$^{-3}$ 
    and the CO column density is fixed  to $N_{\rm CO}$ = 5 $\times$ 10$^{17}$~cm$^{-2}$.
    The solid black lines show the constant [C \emissiontype{I}]~(1--0)/CO~(1--0) line ratios, 
    and the dashed black lines show the constant CO~(3--2)/CO~(1--0) line ratios.
\label{fig:f7}}
\end{figure}

In Figure~\ref{fig:f7}, we show the results of RADEX calculation 
when the H$_{2}$ density is fixed to 
$n_{\rm H_{2}}$ = [0.5, 1.0, 5.0] $\times$ 10$^{4}$\,cm$^{-3}$. 
These fixed densities are similar to or higher than the mean gas densities derived by \citet{Sliwa2017}. 
Their radiative transfer analysis yields the mean gas densities of 10$^{2.54}$--10$^{3.76}$~cm$^{-3}$ 
and the mean kinetic temperatures of 105--240~K. 
Relatively high densities ($\geq 5 \times 10^4$ cm$^{-3}$) are required 
for reproducing high CO (3-2)/CO (1-0) line ratios at the kinetic temperatures of $<$ 300~K. 
In addition, \citet{Israel2015} found that the interstellar medium in LIRGs is dominated by dense 
($n_{\rm H2}$ = 10$^{4}$--10$^{5}$~cm$^{-3}$) and moderately warm ($T_{\rm kin} \sim$ 30~K) gas clouds. 
Considering these results, the density range adopted for our RADEX calculations 
is likely appropriate for typical gas conditions 
in IR-bright galaxies, including Arp~220.

The [C \emissiontype{I}]/CO line ratio decreases with increasing $R_{\rm CO}$ 
(or $T_{\rm kin}$) at the fixed C \emissiontype{I}/CO abundance ratio, 
except under kinetic temperature below 10~K. 
For example, in the case where $n_{\rm H_{2}} = 1 \times 10^{4}$~cm$^{-3}$ 
and $N_{\rm CI}$/$N_{\rm CO}$ = 1.0 (Figure~\ref{fig:f7} (middle)), 
the [C \emissiontype{I}]/CO line ratio is $\sim$0.26 at $R_{\rm CO}$ = 0.8 
and decreases to $\sim$0.19 at $R_{\rm CO}$ = 1.2. 
This is different from the observed trend (Figure~\ref{fig:f6}). 
We also check the modeled line ratios 
when the kinetic temperature is fixed to $T_{\rm k}$ = [75, 100, 125]~K and the H$_{2}$ density is changed. 
In this case, the [C \emissiontype{I}]/CO line ratio does not change significantly 
with increasing $R_{\rm CO}$ (or $n_{\rm H_{2}}$) at the fixed C \emissiontype{I}/CO abundance ratio. 
According to our RADEX calculation, 
the observed trend is not fully explained by changing only the temperature or gas density.

\begin{figure}[htbp]
    \begin{center}
        \includegraphics[width=0.32\textwidth]{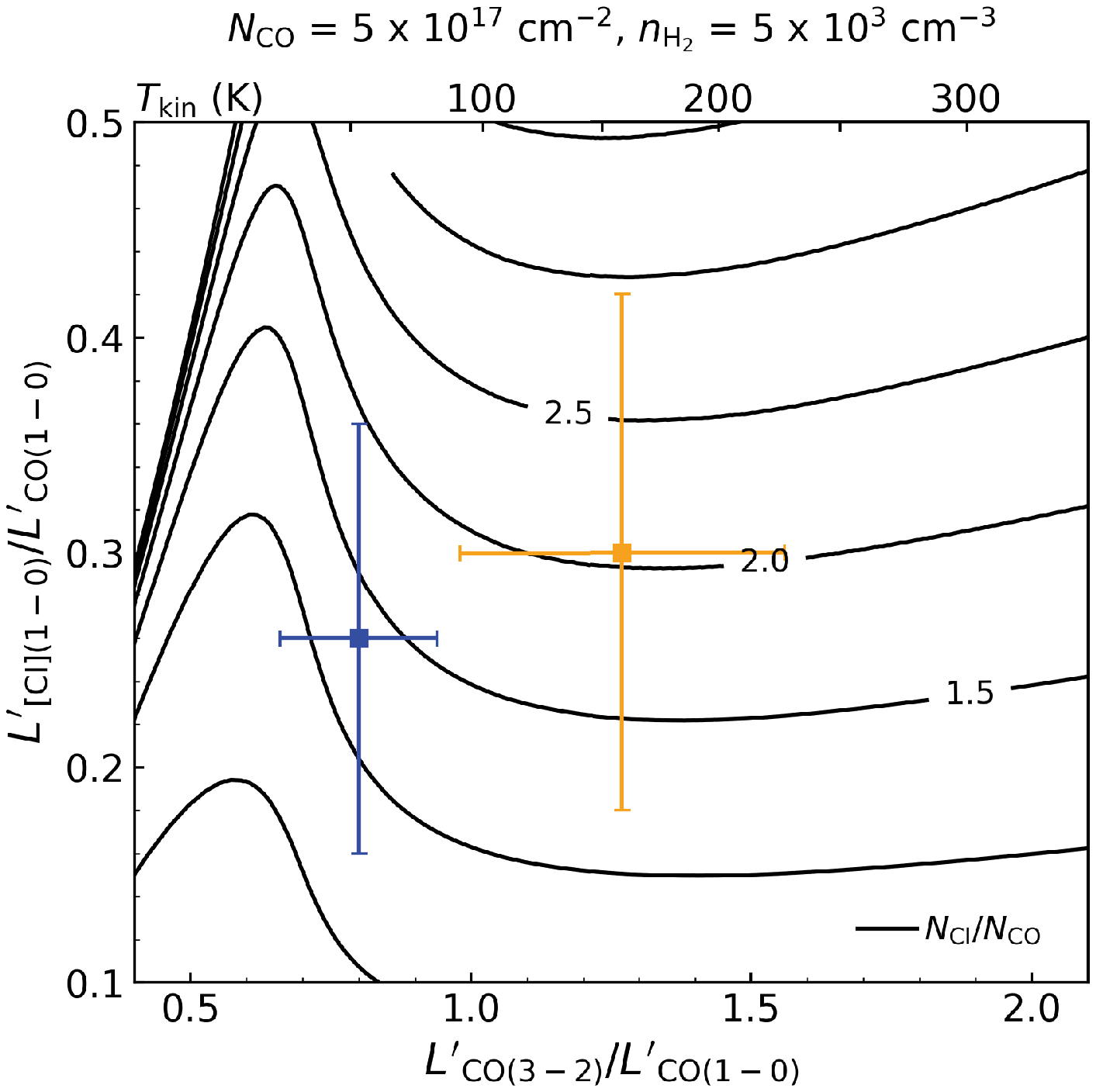}
        \includegraphics[width=0.32\textwidth]{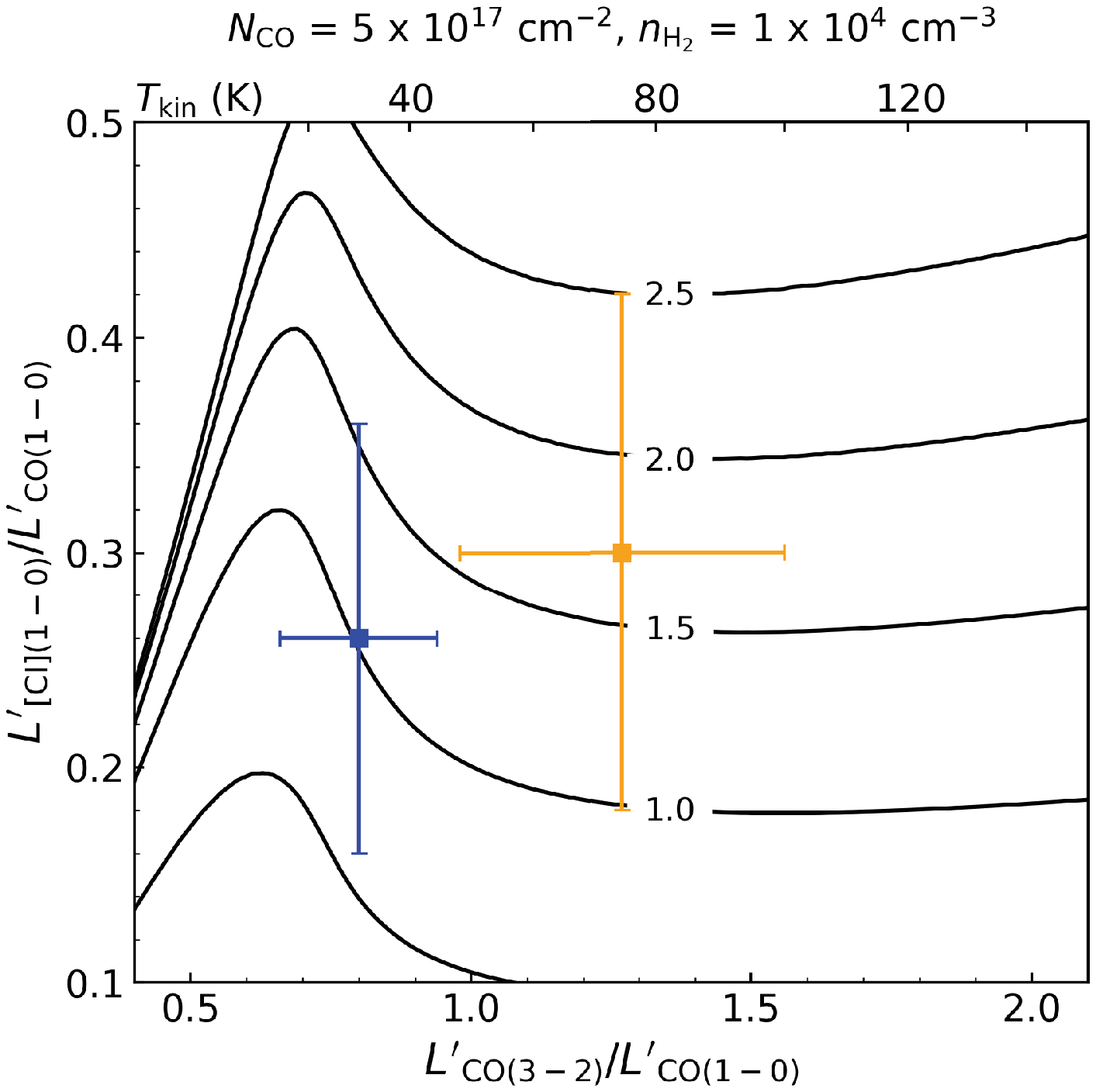}
        \includegraphics[width=0.32\textwidth]{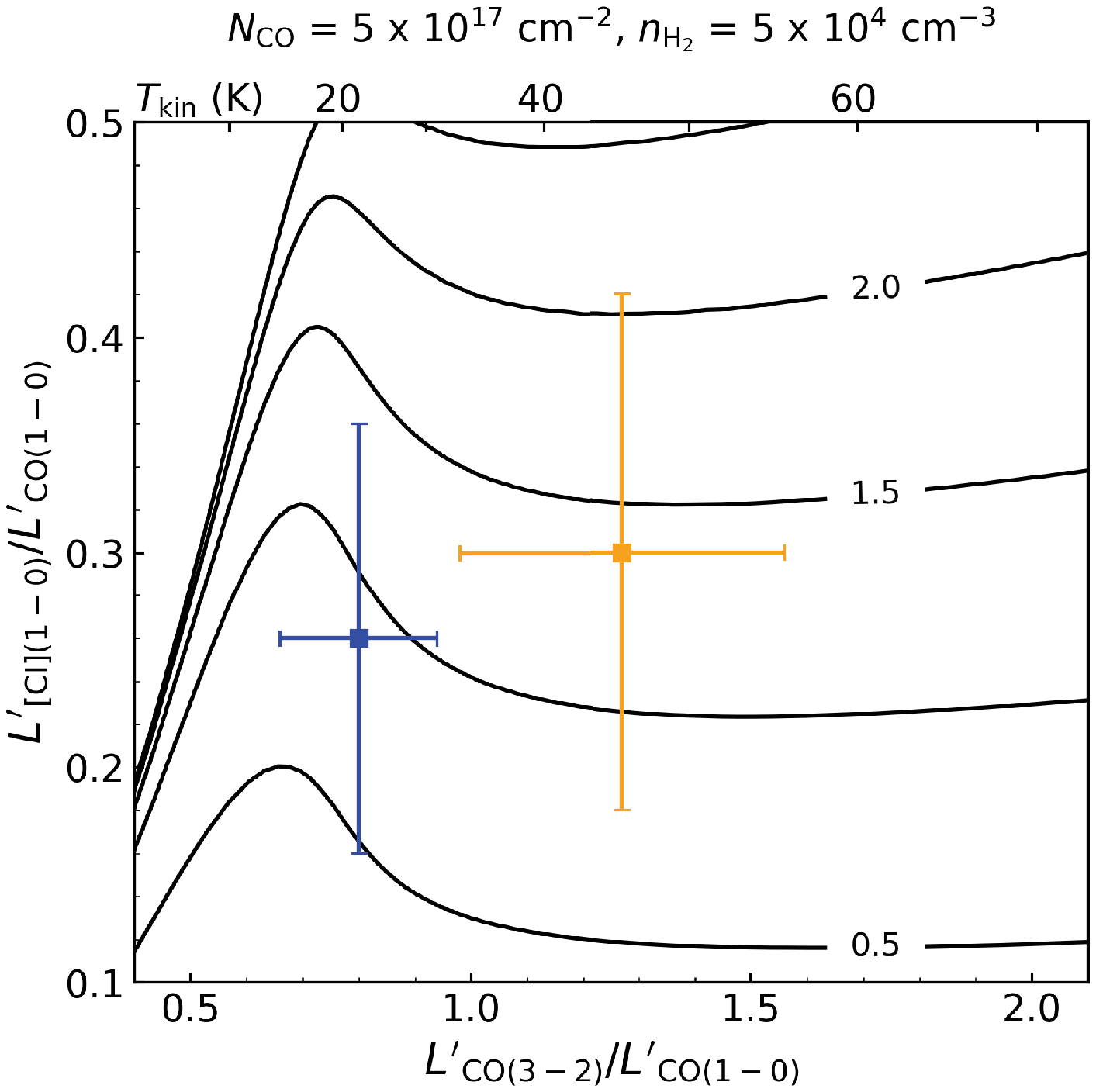}        
    \end{center}
    \caption{
    Plot of the [C \emissiontype{I}]~(1--0)/CO~(1--0) luminosity ratio vs. the CO~(3--2)/CO~(1--0) luminosity ratio 
    for comparison between the observed and modeled data. 
    The orange and blue squares show the representative parameters of Group~1 
    (CO~(3--2)/CO(1--0) $\geq$ 1) and Group~2 (CO~(3--2)/CO~(1--0) $<$ 1), respectively. 
    These parameters are calculated from the observations. 
    The black lines show the constant C \emissiontype{I}/CO abundance ratios 
    calculated from the RADEX analysis when the H$_{2}$ density is fixed to 
    $n_{\rm H_{2}}$ = [0.5, 1.0, 5.0] $\times$ 10$^{4}$~cm$^{-3}$ and 
    the CO column density is fixed to $N_{\rm CO}$ = 5 $\times$ 10$^{17}$~cm$^{-2}$.
\label{fig:f8}}
\end{figure}

Furthermore, we compare the RADEX results with the observational data. 
For simplicity, we divide the observational data points (pixels) into two groups by $R_{\rm CO}$. 
Group~1 has $R_{\rm CO} \geq 1$, and Group~2 has $R_{\rm CO} < 1$. 
Then we compute the mean $L'_{\rm [CI](1-0)}/L'_{\rm CO(1-0)}$ and $R_{\rm CO}$ of each group. 
Group~1 has $L'_{\rm [CI](1-0)}/L'_{\rm CO(1-0)}$ = 0.30 $\pm$ 0.12 and $R_{\rm CO}$ = 1.27 $\pm$ 0.29, 
and Group~2 has $L'_{\rm [CI](1-0)}/L'_{\rm CO(1-0)}$ = 0.26 $\pm$ 0.10 and $R_{\rm CO}$ = 0.80 $\pm$ 0.14. 
In Figure~\ref{fig:f8}, we show the parameter sets ( $L'_{\rm [CI](1-0)}/L'_{\rm CO(1-0)}$ and $R_{\rm CO}$) of Group~1 
and Group~2 by the orange and blue squares, respectively, 
and plot lines to show the constant C \emissiontype{I}/CO abundance ratios calculated from the RADEX analysis 
when the H$_{2}$ density is fixed to $n_{\rm H_{2}}$ = [0.5, 1.0, 5.0] $\times$ 10$^{4}$\,cm$^{-3}$. 
We find that the C \emissiontype{I}/CO abundance ratio at Group~1 is similar to or a few times higher than that at Group~2. 
For example, in the case of $n_{\rm H_{2}} = 1 \times 10^{4}$~cm$^{-3}$ (Figure~\ref{fig:f8} (middle)), 
the C \emissiontype{I}/CO abundance ranges from 0.69 to 2.00 for Group~1, 
and it ranges from 0.18 to 1.32 for Group~2. 
While the C \emissiontype{I}/CO abundance ranges partially overlap between the two groups, 
an elevated abundance ratio is likely to be required 
to explain the observed high $L'_{\rm [CI](1-0)}/L'_{\rm CO(1-0)}$ in regions with high $R_{\rm CO}$. 
We obtain similar results when calculating the abundance ratio by changing the $H_{2}$ density. 
The study of the central region of the starburst galaxy NGC~1808 reaches 
a similar conclusion that $L'_{\rm [CI](1-0)}/L'_{\rm CO(1-0)}$ is enhanced 
owing to high excitation and atomic carbon abundance \citep{Salak2019}. 
We also check the RADEX results when the parameters ($N_{\rm co}$ or d$V$) are fixed to different values. 
For example, we adopt $N_{\rm CO}$ = [0.1, 0.5, 1.0] $\times 10^{18}$ cm$^{-2}$. 
While the C \emissiontype{I}/CO abundance ratios for Group~1 are on average higher than those for Group~2 in all three cases, 
the difference between Group~1 and Group~2 decreases with decreasing $N_{\rm CO}$. 
The same trend is seen when d$V$ is increased, 
which is a natural consequence of the RADEX calculation as it depends on $N_{\rm co}$/d$V$. 
In this study, assuming a constant $N_{\rm CO}$, we have performed the radiative transfer analysis. 
If $N_{\rm CO}$ is significantly changing from one region to another in the science target, 
the cause of various $L'_{\rm [CI](1-0)}$/$L'_{\rm CO(1-0)}$ cannot be straightforward to interpret. 
Under the assumption that $N_{\rm CO}$ is constant in the observed region, 
the comparison between the observational and modeled data suggests that 
the C \emissiontype{I}/CO abundance ratio varies at $\sim$100 pc scale in Arp~220.

\subsection{High line ratio in the redshifted component}
\begin{figure}[htbp]
    \begin{center}
        \includegraphics[width=0.32\textwidth]{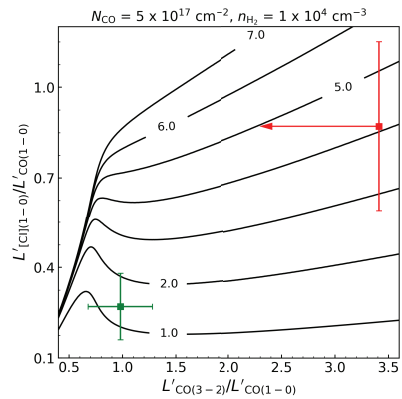}
    \end{center}
    \caption{
    The same as Figure~\ref{fig:f8}, but the symbols are different.
    The red and green squares show the average line ratios 
    for the redshifted component and the main body of Arp~220, respectively. 
    The black lines show the constant C \emissiontype{I}/CO abundance ratios 
    calculated from the RADEX analysis when the CO column density is fixed 
    to $N_{\rm CO}$ = 5 $\times$ 10$^{17}$~cm$^{-2}$ and 
    the H$_{2}$ density is fixed to $n_{\rm H_{2}}$ = 1.0 $\times$ 10$^{4}$~cm$^{-3}$.
\label{fig:f9}}
\end{figure}

We calculate the [C \emissiontype{I}]/CO luminosity ratios of the high-velocity components 
using pixels in which both [C \emissiontype{I}]~(1--0) and CO~(1--0) flux densities are above the 3$\sigma$ levels. 
The average $L'_{\rm [CI](1-0)}/L'_{\rm CO(1-0)}$ of the redshifted component is 0.87 $\pm$ 0.28, 
which is four times higher than 
the average $L'_{\rm [CI](1-0)}/L'_{\rm CO(1-0)}$ of Arp~220 (0.22 $\pm$ 0.04) (see Figure~\ref{fig:f5}). 
The average $L'_{\rm [CI](1-0)}/L'_{\rm CO(1-0)}$ of the blueshifted component is 0.41 $\pm$ 0.11, 
but only five pixels meet the threshold. 
High [C \emissiontype{I}]/CO line ratios have been found in the root point of the bipolar outflow 
of the NGC~253 ($\sim$0.4--0.6; \cite{Krips2016}) 
and the bipolar outflow of the Galactic molecular cloud G5.2-0.74N ($\sim$0.4; \cite{Little1998}). 
While we cannot rule out the possibility that the [C \emissiontype{I}]~(1--0) and CO~(1--0) emission 
traces different components, as mentioned in \S3.3, 
we suggest that the high [C \emissiontype{I}]/CO line ratios are caused by 
an elevated C \emissiontype{I}/CO abundance ratio based on our RADEX analysis (Figure~\ref{fig:f9}). 
Since the CO~(3--2) spectra of the redshifted component are affected 
by contamination from other molecular lines, possibly H$^{13}$CN~(4--3), 
we calculate an upper limit for $R_{\rm CO}$ in the redshifted component 
and plot it with $L'_{\rm [CI](1-0)}/L'_{\rm CO(1-0)}$ of the redshifted component in Figure~\ref{fig:f9}. 
The high C \emissiontype{I}/CO abundance ratio is required 
for the [C \emissiontype{I}]/CO luminosity ratios measured in the redshifted component, 
regardless of the value of $R_{\rm CO}$.

The possibility of CO-poor/C \emissiontype{I}-rich molecular gas in outflows 
is predicted by the theoretical study \citep{Papadopoulos2018}. 
Low-density, gravitationally-unbound molecular gas is expected to be in outflows 
because of Kelvin-Helmholtz instabilities and shear acting on the envelopes of denser clouds. 
Such a gas phase could be invisible in CO due to the destruction of CO induced by the CRs, 
which have been identified as a potentially more effective agent of CO-destruction than far-UV photons 
(e.g., \cite{Bialy2015, Bisbas2015, Papadopoulos2018}). 
\citet{Varenius2016} found an elongated feature extending 0\farcs9 from the western nucleus 
in the radio continuum maps at 150~MHz, 1.4~GHz, and 6~GHz, indicating the presence of outflow. 
This elongated feature requires the shock-acceleration of CRs in the outflow. 
In addition, the $\gamma$-ray observations suggest the presence of an AGN, 
providing the extra CRs, in the western nucleus \citep{Yoast-Hull2017}. 
In the high-velocity components of Arp 220, CRs could help enrich C \emissiontype{I} by destroying CO.

Another possibility of the high $L'_{\rm [CI](1-0)}/L'_{\rm CO(1-0)}$ is 
the CO dissociation by mechanical perturbation such as shocks and the strong radiation field. 
The dissociation of C-bearing molecules such as CO can significantly enhance the C abundance 
in active environments where shocks occur and/or strong UV and X-ray radiation fields 
associated with intense star formation and an AGN (e.g., \cite{Meijerink2007}, \cite{Meijerink2011}, \cite{Tanaka2011}). 
\citet{Krips2016} found that the [C \emissiontype{I}]~(1--0) line emission is enhanced 
compared to CO~(1--0) in the region of NGC~253 
where outflows or shocks are important dynamical players, 
suggesting that shocks affect the [C \emissiontype{I}]/CO ratio.

\section{Summary}
We present the new [C \emissiontype{I}]~(1--0) map of Arp~220 and the uv-matched CO~(1--0) map as well. 
The distributions and kinematics of the [C \emissiontype{I}]~(1--0) and CO~(1--0) emission are similar overall, 
but the [C \emissiontype{I}]~(1--0) velocity dispersion is 1.4 times higher than the CO~(1--0) in the western nucleus.

The [C \emissiontype{I}]/CO luminosity ratio integrated over the system is 0.22 $\pm$ 0.04, 
which is comparable to the previous measurements of 20 U/LIRGs, 
and the [C \emissiontype{I}]~(1--0) and CO~(1--0) luminosities of Arp~220 are consistent with the empirical linear relation 
between $L'_{\rm [CI](1-0)}$ and $L'_{\rm CO(1-0)}$ for the unresolved ULIRG sample. 
However, when measuring the [C \emissiontype{I}]~(1--0) and CO~(1--0) luminosities per pixel 
(1~pix = 57~pc $\times$ 57~pc, $\Delta v$ = 5~km~s$^{-1}$), we find a sublinear relation between them, 
indicating that the [C \emissiontype{I}]/CO luminosity ratios are not constant in Arp~220.

We investigate the dependence of the [C \emissiontype{I}]/CO luminosity ratio on $R_{\rm CO}$. 
The average $L'_{\rm [CI](1-0)}/L'_{\rm CO(1-0)}$ is almost constant 
up to $R_{\rm CO}$ = 1.0 and then increases with $R_{\rm CO}$. 
We perform non-LTE analysis using the radiative transfer code RADEX 
to check whether this trend can be reproduced by changing the physical properties of the gas or the abundance ratio. 
As a result, a high abundance ratio is required to explain 
the enhanced $L'_{\rm [CI](1-0)}/L'_{\rm CO(1-0)}$ in regions with $R_{\rm CO} > 1$. 
This suggests that the C \emissiontype{I}/CO abundance ratio varies at $\sim$100 pc scale within the system of Arp~220.

Finally, the atomic carbon was detected in the high-velocity components, which are likely to be the outflow. 
The average $L'_{\rm [CI](1-0)}/L'_{\rm CO(1-0)}$ of the redshifted component is 0.87 $\pm$ 0.28, 
which is four times higher than the average $L'_{\rm [CI](1-0)}/L'_{\rm CO(1-0)}$ of Arp~220. 
The theoretical work on the CR-driven astrochemistry supports 
such a gas phase of C \emissiontype{I}-rich and CO-poor. 
CRs could help enrich CI by destroying CO in the outflowing gas.

\begin{ack}
We thank Qian Jiao for providing the data of local galaxies published in \citet{Jiao2019}.
We also thank Kouichiro Nakanishi for advising us about the data reduction.
J.U. was supported by the ALMA Japan Research Grant of NAOJ ALMA Project, NAOJ-ALMA-258.
D.I. is supported by JSPS KAKENHI Grant Number JP18H03725.

This paper makes use of the following ALMA data: 
ADS/JAO.ALMA\#2013.1.00368.S, 
ADS/JAO.ALMA\#2015.1.00113.S,
ADS/JAO.ALMA\#2017.1.00042.S.
ALMA is a partnership of ESO (representing its member states), 
NSF (USA) and NINS (Japan), together with NRC (Canada), 
MOST and ASIAA (Taiwan), and KASI (Republic of Korea), 
in cooperation with the Republic of Chile. 
The Joint ALMA Observatory is operated by ESO, AUI/NRAO and NAOJ.

Data analysis was carried out on the Multi-wavelength Data Analysis System 
operated by the Astronomy Data Center (ADC), National Astronomical Observatory of Japan.
\end{ack}

\clearpage
\onecolumn
\begin{table}
    \tbl{The [C \emissiontype{I}] and CO measurements in the central 5\arcsec region of Arp~220}{
    \begin{tabular}{ll}
    \hline
    $S_{\rm [CI]}\Delta v$ & 760 $\pm$ 150 Jy~km~s$^{-1}$\\
    $S_{\rm CO}\Delta v$ & 190 $\pm$ 10 Jy~km~s$^{-1}$\\
    $L'_{\rm [CI]}$ & (6.6 $\pm$ 1.3) $\times$ 10$^{8}$ K~km~s$^{-1}$~pc$^{-2}$\\
    $L'_{\rm CO}$ & (3.0 $\pm$ 0.2) $\times$ 10$^{9}$ K~km~s$^{-1}$~pc$^{-2}$ \\
    $L'_{\rm [CI]}/L^{'}_{\rm CO}$ & 0.22 $\pm$ 0.04\\
    \hline
    \end{tabular}
    }\label{tab:t1}
\end{table}

\begin{table}
    \tbl{The [C \emissiontype{I}] and CO measurements of the high-velocity components}{
    \begin{tabular}{cccccc}
    \hline
    Component & $S_{\rm [CI]}\Delta v$ & $S_{\rm CO}\Delta v$ & $M_{\rm CI}$ & $M_{\rm CO}$ & $M_{\rm CI}$/$M_{\rm CO}$\\
    & (Jy~km~s$^{-1}$) & (Jy~km~s$^{-1}$) & ($M_{\solar}$) & ($M_{\solar}$) &\\
    \hline
    Redshifted component & 5.3 $\pm$ 1.1 & 0.25 $\pm$ 0.03 & $(1.7 \pm 0.4) \times 10^{7}$ & $(3.2 \pm 0.4) \times 10^{6}$ & 5.3 $\pm$ 1.3\\
    Blueshifted component & 1.0 $\pm$ 0.2 & 0.33 $\pm$ 0.03 & $(3.2 \pm 0.6) \times 10^{6}$ & $(4.2 \pm 0.4) \times 10^{6}$ & 0.76 $\pm$ 0.17\\ 
    \hline
    \end{tabular}
    }\label{tab:t2}
    \begin{tabnote}
    Columns~2 and 3: The integrated [C \emissiontype{I}]~(1--0) and CO~(1--0) line flux density 
    measured within $0\farcs8 \times 1\farcs0$ from the CO~(1--0) peak of each high-velocity component. 
    Columns~4 and 5: The molecular gas mass estimated from the [C \emissiontype{I}]~(1--0) and CO~(1--0) line flux density 
    by applying $X_{\rm CI} = 7.5 \times 10^{-5}$ and $\alpha_{\rm CO}$ = 0.8 $M_{\solar}$ pc$^{-2}$ (K~km~s$^{-1}$)$^{-1}$.
    \end{tabnote}
\end{table}

\clearpage
\appendix
\section*{Channel maps of the [C \emissiontype{I}]~(1--0)/CO~(1--0) luminosity ratio}

We present a figure with channel maps of the [C \emissiontype{I}]/CO luminosity. 
We rebinned three channels to reduce the number of channels for plotting. 
The final velocity resolution is 15 km~s$^{-1}$.

\begin{figure}[htbp]
    \begin{center}
        \includegraphics[width=0.7\textwidth]{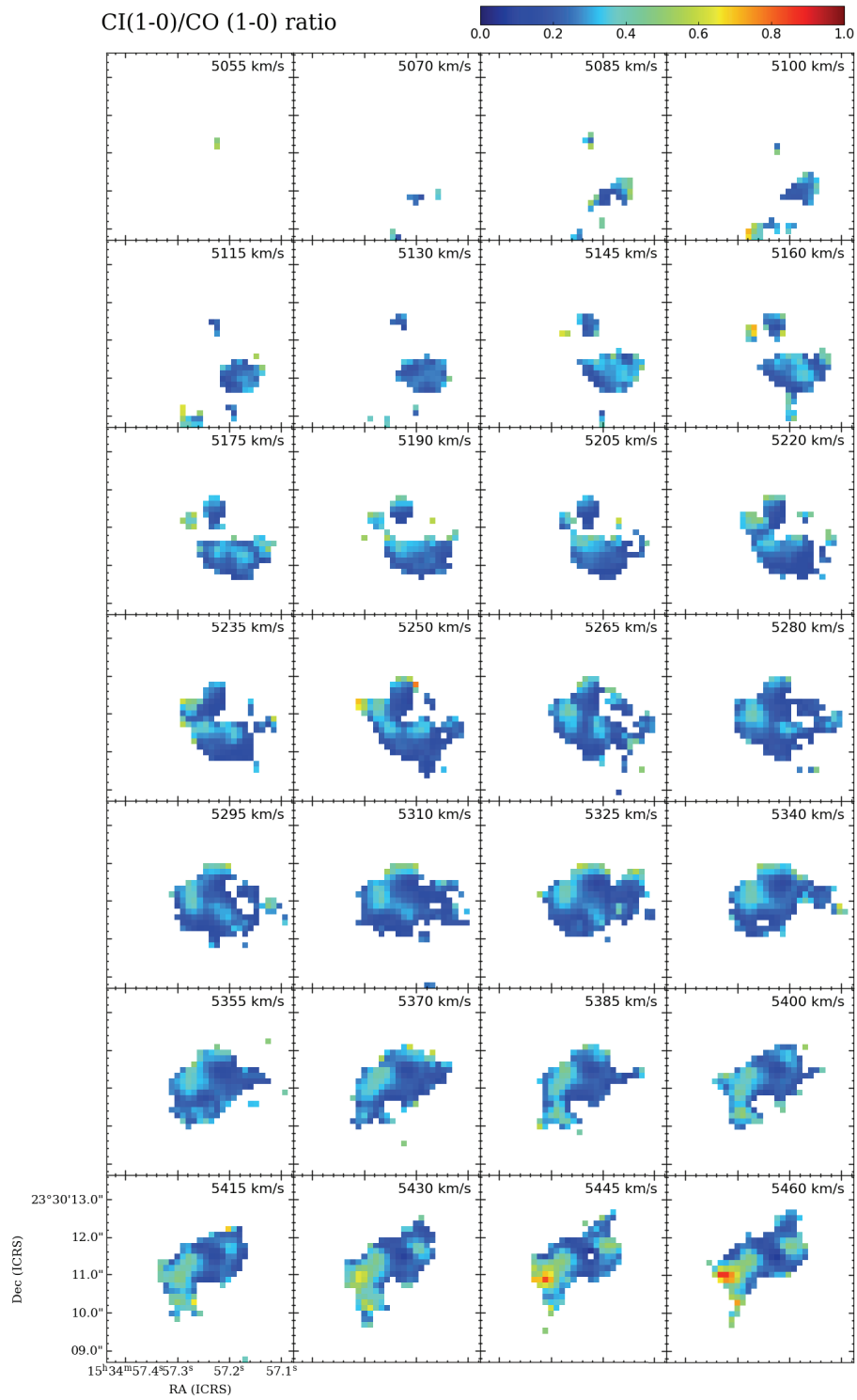}
    \end{center}
    \caption{
    Channel maps of the [C \emissiontype{I}]~(1--0)/CO~(1--0) luminosity ratio. 
    The velocity resolution is 15 km~s$^{-1}$.
    \label{fig:f.a1}}
\end{figure}

\addtocounter{figure}{-1}
\begin{figure}[htbp]
    \begin{center}
        \includegraphics[width=0.7\textwidth]{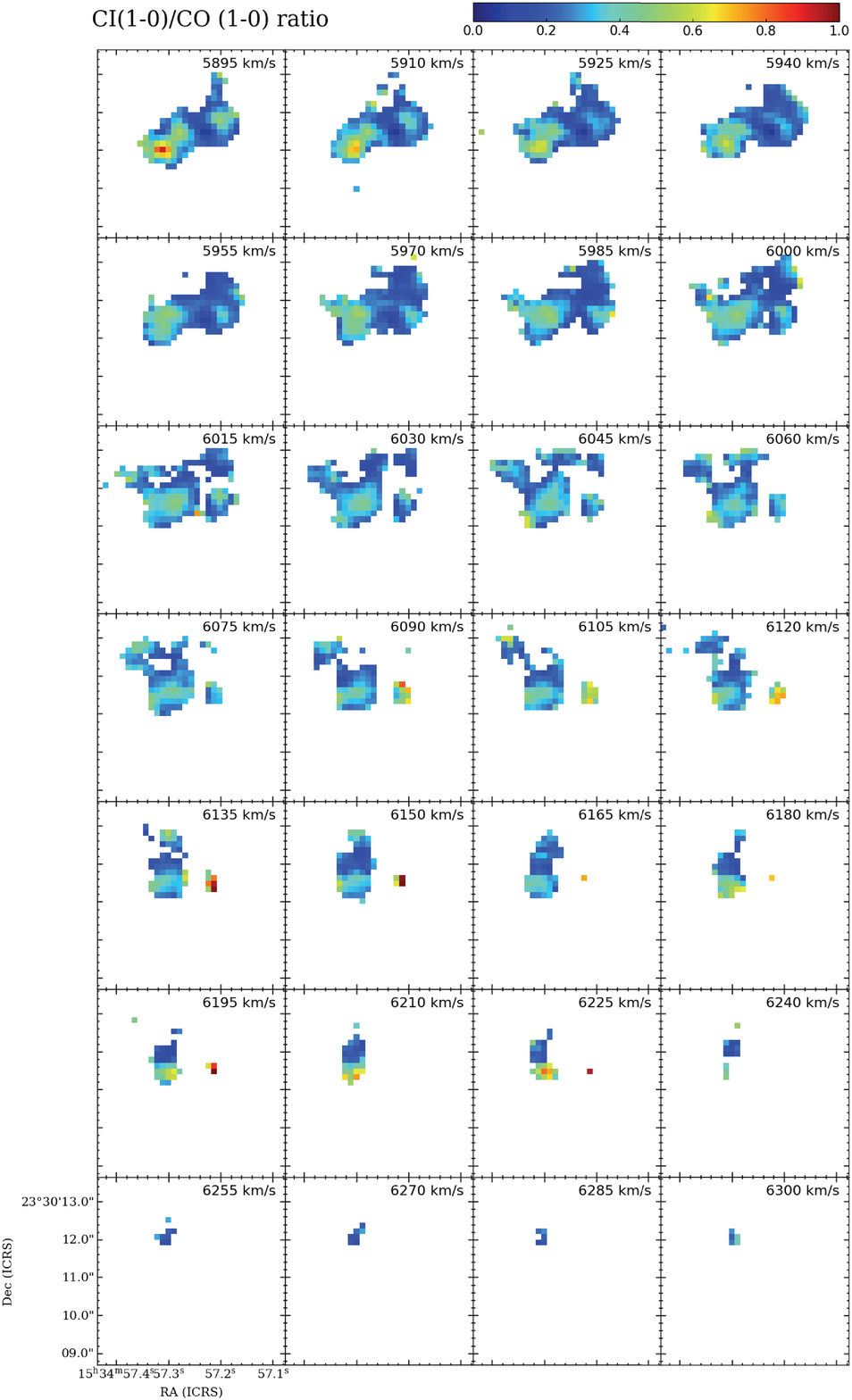}
    \end{center}
    \caption{
    Continued.
    \label{fig:f.a1}}
\end{figure}

\end{document}